\documentstyle[aps,preprint]{revtex}

\def\[{\left\lbrack}
\def\]{\right\rbrack}

\def\({\left(}
\def\){\right)}
\def\ih{\'\i}

\newcommand{\be}{\begin{equation}}
\newcommand{\ee}{\end{equation}}
\newcommand{\ea}{\end{eqnarray}}
\newcommand{\ba}{\begin{eqnarray}}

\begin{document}

\title{Gauging by symmetries}

\author{J. Ananias Neto\thanks{jorge@fisica.ufjf.br}, C. Neves\thanks{cneves@fisica.ufjf.br} and W. Oliveira\thanks{wilson@fisica.ufjf.br}}
\address{Departamento de F\ih sica, ICE,\\ Universidade Federal de Juiz de Fora, 36036-330, Juiz de Fora, MG, Brazil }

\maketitle

\pagestyle{myheadings}
\markright{J. Ananias Neto, C. Neves and W. Oliveira, `Gauging by symmetries'}

\begin{abstract}
We propose a new procedure to embed second class systems by introducing Wess-Zumino (WZ) fields in order to unveil hidden symmetries existent in the models. This formalism is based on the direct imposition that the new Hamiltonian must be invariant by gauge-symmetry transformations. An interesting feature in this approach is the possibility to find a representation for the WZ fields in a convenient way, which leads to preserve the gauge symmetry in the original phase space.  Consequently, the gauge-invariant Hamiltonian can be written only in terms of the original phase-space variables. In this situation, the WZ variables are only auxiliary tools that permit to reveal the hidden symmetries present in the original second class model. We apply this formalism to important physical models: the reduced-SU(2) Skyrme model, the Chern-Simons-Proca quantum mechanics and the chiral bosons field theory. In all these systems, the gauge-invariant Hamiltonians are derived in a very simple way.
\end{abstract}

\newpage

\setlength{\baselineskip}{20pt}  
\renewcommand{\baselinestretch}{2}

\section{Introduction}

It is well known that through symmetries, important properties present on physical systems might be investigated in a more general way. In view of this, many works\cite{BFFT,embed,JW} have embedded second class systems into first class ones by enlarging the phase-space with the introduction of Wess-Zumino (WZ) variables.  At the same time, an alternative gauge-invariant approach\cite{MR,Vyt} has been proposed, which considers part of the total second class constraints as gauge fixing terms, while the remaining ones form a subset that satisfies a first class algebra. This formalism has an elegant property that does not extend the phase-space with extra variables. 

The main feature of this paper is to propose an alternative scheme to embed noninvariant models and, 
in some cases, to extract hidden symmetries existent
in those models. This new approach mixes the WZ and projection concepts idealized in Ref. \cite{FS} and Refs.\cite{MR,Vyt}, respectively. We will extend the initial noninvariant gauge Hamiltonian with the introduction of an arbitrary function $ G $ written in terms of the original phase space and WZ variables, as suggested by Faddeev\cite{FS}. Afterward, the extended Hamiltonian is constructed  such as it must satisfy the variational condition, $\delta H = 0$, i.e.,  the new Hamiltonian must be invariant by gauge-symmetry transformations. Here, it is opportune to mention that symmetries, obtained in the other constrained conversion formalisms\cite{BFFT,embed,JW}, appear as a consequence of the first class conversion mechanism. We will see that the possibility of choosing particular symmetries for the WZ terms leads to the considerable simplifications in the determination of the gauge invariant Hamiltonian.  Further, we show that the WZ terms can be determined in some cases, and consequently, the invariant Hamiltonian can be written only as function of the original phase-space variables. It is important to mention that the WZ fields, 
in these cases, are auxiliary variables that permit to reveal the hidden symmetries present in the original system.

In order to clarify the exposition of the subject, the paper is organized as follows. In Sec. II, the formalism, which it will be called  as ``variational gauge-invariant formalism'',  will be presented in detail. In Sec. III, we 
will apply this formalism on some important physical systems in order to unveil the hidden symmetries.  We begin considering to study the SU(2) Skyrme model\cite{Skyrme}, which is an effective field theory to describe hadrons physics. Nappi {\it et al.}\cite{ANW}, using collective coordinates, reduced the SU(2) Skyrme model to a nonlinear quantum mechanical model depending explicitly on the time-dependent collective variables, which satisfies a spherical constraint. Due to the nonlinear structure of the Skyrme model, here we consider this second class constrained system as a nontrivial application of our gauge-invariant formalism. Afterward, the Chern-Simons-Proca (CSP) model\cite{DJT,niemi} is considered. It is a quantum mechanical system obtained from the Abelian Chern-Simons theory, which has the gauge field modified by adding a Proca mass term. This modification leads to a significant perturbation of the Chern-Simons action in the infrared limit, which could have physically relevant consequences, for example, in the quantum Hall effect or high-temperature superconductivity. In Ref. \cite{niemi}, the authors investigate the infrared limit of the Abelian Chern-Simons-Proca theory and found that this limit can be described by two {\it a priori} different topological quantum mechanical model\cite{DJT}. This model exhibits a noncommutative geometry and an important result of our gauge-invariant conversion formalism is the obtainment of a commutative version of the Chern-Simons-Proca quantum mechanical model. To finish, we consider the two dimensional self-dual boson theory proposed by Siegel\cite{siegel} many years ago. However, this description could not be quantized because it is anomalous at the quantum level. Many authors have attempted to solve this problem following different strategies, however, it was not achieved until the Floreanini and Jackiw's paper \cite{FJ}, where the self-dual fields were quantized in a consistent way just basing the process on an unconventional Poincar\'e symmetry. In this section, we will see that the WZ auxiliary field is essential to derive the first class version of the two dimensional chiral-boson field theory.
In Sec. IV, the last section, we will discuss our findings together with our final comments and conclusions.

\section{General formalism}
In this section, we present a sketch of the variational gauge-invariant formalism. To this end, a general second class constrained mechanical system is considered to study. This system has the dynamic governed by a Lagrangian ${\cal L}(q_i,\dot q_i,t)$ ( $i=1,2,\dots,N$), where $q_i$ and $\dot q_i$ are the space and velocities variables, respectively, with a set of second class constraints ($T_a(q_i,p_i),\; a=1,2,\dots, M,\,\,M<N$, where $p_i$ is the canonical momentum conjugated to the variable $q_i$). Notice that this consideration does not lead to lost generality or physical content. In order to systematize this formalism, we separate the development following two steps. The first one is to consider the set of second class constraints and consecutive split up it in two subset, where one is chosen to construct the $\; R\, (R\leq M)\;$ symmetry generators with the auxiliary WZ variables $(\theta_\alpha,\pi_{\theta_\alpha})$, while the other one is considered as being the gauge-fixing terms which are discarded by the variational gauge-invariant formalism. Here, we would like to mention that the number of first class symmetry generators must be chosen such that the resulting first class theory have the same physical degrees of freedom of the original second class model. This equality, certainly, indicates the equivalence of the converted first class theory and the initial second class model. In general, given the initial $2N$ second class constraints, we must select $N$ constraints (except in the chiral bosons field theory as we will see) to be the generators of the symmetry. In the second step, an arbitrary function $ G $ dependent on the original phase space variables $(q_i,p_i)$ and WZ fields, $\theta_\alpha$,  is added in the canonical Hamiltonian with the purpose to  construct a gauge-invariant Hamiltonian. We impose that this new Hamiltonian, $\tilde H$, must be invariant by gauge-symmetry transformations, using for this, the $\; R\;$ symmetry generators(see Eq.(\ref{gstrans})), initially defined.  Consequently, this procedure leads to the complete determination of the arbitrary function. 

Let us to start considering the following set of second class constraints

\begin{eqnarray}
\label{cons}
T_a(q_i,p_i)\approx 0, \,\,\, \,{\text with \,\,}\,\, a=1,2,\dots,M \, 
\end{eqnarray}
obtained through the iterative Dirac's procedure. Note that the Dirac's matrix, given by
\begin{eqnarray}
D_{ab} = \lbrace T_a, T_b\rbrace,
\end{eqnarray}
it is a nonsingular matrix. In this scenario, the system has its dynamic described by the canonical Hamiltonian $(H_c\equiv H_c(q_i,p_i))$ together with
the Dirac's brackets among the phase space variables. Notice that in this context, the gauge symmetry is hidden. Thus, our purpose is to reveal this symmetry, in principle, enlarging the original set of phase space variables by introducing WZ fields $\theta_\alpha$. At the end, in some cases, these WZ fields can be determined in a convenient way in order to unveil the gauge symmetry in the original set of phase space variables.

The variational gauge-invariant formalism begins forging the algebraic form of the generators of the symmetry with the WZ terms as

\begin{equation}
\label{fgenerator}
\tilde T_\alpha = C_\alpha T_\alpha 
+ D_\alpha T_{\theta_\alpha}, \,\,\,\,{\text with\,}\,\,\, \alpha=1,2,\dots, R
\end{equation}
where $T_{\theta_\alpha}$ is a function of the WZ variables  $(\theta_\alpha, \pi_{\theta_\alpha})$ and $C_\alpha$ and $D_\alpha$ are constants. Note that there is not sum in $\alpha$ indices in Eq.(\ref{fgenerator}). Further, the first class algebra among these generators, Eq.(\ref{fgenerator}), must be obeyed, namely,

\begin{eqnarray}
\label{brac}
\lbrace{\tilde T}_\alpha, {\tilde T}_\beta \rbrace &=& 0,
\end{eqnarray}
which restricts the values assumed by the constants $C_\alpha$ and $D_\alpha$.
  
As ${\tilde T}_\alpha$ is the symmetry generator, the infinitesimal gauge transformations can be computed as well, namely,

\begin{eqnarray}
\label{GI}
\delta q_i &=&\varepsilon_\alpha \lbrace q_i,{\tilde T}_\alpha\rbrace = \varepsilon_\alpha C_\alpha\lbrace q_i,T_\alpha\rbrace,\nonumber\\
\delta p_i &=& \varepsilon_\alpha\lbrace p_i,{\tilde T}_\alpha\rbrace = \varepsilon_\alpha C_\alpha\lbrace p_i,T_\alpha\rbrace,\\
\delta \theta_\alpha &=& \varepsilon_\alpha\lbrace \theta_\alpha,{\tilde T}_\alpha\rbrace = \varepsilon_\alpha D_\alpha\lbrace \theta_\alpha, T_{\theta_\alpha}\rbrace.\nonumber
\end{eqnarray}

The gauge-invariant Hamiltonian, $\tilde{H}$, is built by adding an arbitrary function $G(q_i,p_i,\theta_\alpha)$ to the canonical 
Hamiltonian $(H_c\equiv H_c(q_i,p_i))$, namely,
\begin{equation}
\label{FHamiltonian}
\tilde{H}= H_c + G(q_i,p_i,\theta_\alpha),
\end{equation}
where the arbitrary function $G(q_i,p_i,\theta_\alpha)$ is expressed as an expansion in order of the WZ fields, which is written as

\begin{equation}
\label{2060}
G(q_i,p_i,\theta_\alpha)=\sum_{n=1}^\infty \sum_{\alpha=1}^R G_{(n)}^\alpha(q_i,p_i)\theta_\alpha^n, \,\, {\text with}\,\,\, R\leq M.
\end{equation}
Note that this function satisfies the following boundary condition

\begin{equation}
\label{2070}
G(q_i,p_i,\theta_\alpha=0) = 0.
\end{equation}
Thus, it is important to remark that the introduction of the arbitrary function $\,G\,$ 
will permit the construction of a gauge-invariant Hamiltonian 
from the canonical second class Hamiltonian.

In order to obtain the arbitrary function, the variational condition 

\begin{equation}
\label{gstrans}
\delta{\tilde H}_\alpha=\lbrace \tilde H,{\tilde T}_\alpha \rbrace =0,
\end{equation}
must be obeyed by the WZ extended Hamiltonian, given in Eq.(\ref{FHamiltonian}). Consequently, for the correction terms $G_{(n)}^\alpha(q_i,p_i)$, present in Eq.(\ref{2060}), we have  the following general equation

\begin{equation}
\label{variational}
\delta \tilde H^\alpha = \delta H_c^\alpha + \sum_{n=1}^\infty \left(\delta G_{(n)}^\alpha(q_i,p_i)\theta^n_\alpha + n G_{(n)}^\alpha(q_i,p_i)\theta^{(n-1)}_\alpha\delta\theta_\alpha\right) = 0,
\end{equation}
which allows us to compute each correction term $G_{(n)}^\alpha(q_i,p_i)$. Again, there is not sum in the $\alpha$ indices in Eq.(\ref{variational}). For linear correction term ($n=1$), we have R relations

\begin{eqnarray}
\label{rdeltaH}
\delta H_c^1 &+& G_{(1)}^1(q_i,p_i) \delta\theta_1 = 0,\nonumber\\
\delta H_c^2 &+& G_{(1)}^2(q_i,p_i) \delta\theta_2 = 0,\nonumber\\
&\vdots &\nonumber\\
\delta H_c^R &+& G_{(1)}^R(q_i,p_i) \delta\theta_R = 0.
\end{eqnarray}
For the quadratic correction term ($n=2$), we get $R$  relations

\begin{eqnarray}
\label{rdeltaH2}
\delta  G_{(1)}^1(q_i,p_i) &+& 2 G_{(2)}^1(q_i,p_i) \delta\theta_1 = 0,\nonumber\\
\delta  G_{(1)}^2(q_i,p_i) &+& 2 G_{(2)}^2(q_i,p_i) \delta\theta_2 = 0,\nonumber\\
&\vdots&\nonumber\\
\delta  G_{(1)}^R(q_i,p_i) &+& 2 G_{(2)}^R(q_i,p_i) \delta\theta_R = 0.
\end{eqnarray}
For $n\geq 2 $,  the general relation is

\begin{equation}
\label{ITER}
\delta G_{(n)}^\alpha(q_i,p_i) + (n+1) G_{(n+1)}^\alpha(q_i,p_i) \delta\theta_\alpha = 0.
\end{equation}
This iterative process is successively repeated until the recursive relations (\ref{ITER}) becomes identically null. It leads to a complete determination of the arbitrary function $G(q_i,p_i,\theta_\alpha)$ and, consequently, a complete determination of the invariant Hamiltonian $\tilde{H}$.  Note that, in our formalism,  the recursion relations (\ref{rdeltaH}), (\ref{rdeltaH2}) and (\ref{ITER})  presuppose that the transformation for $\delta\theta$ must be linear, i.e., independent of $\theta$, since powers of $\theta$ are being compared. Here, it is also opportune to mention that, due to the Eq. (\ref{fgenerator}), $\delta G_{(n)}^\alpha$ does not involve $\theta$ itself.

In some cases, there is the possibility that second class systems can not be described by a gauge invariant Hamiltonian $\tilde{H}$ expressed with a finite sum of WZ terms. In this case, due to the relations (\ref{rdeltaH}), (\ref{rdeltaH2}) and (\ref{ITER}) , $\tilde{H}$, Eq.(\ref{FHamiltonian}), can be written as
\begin{eqnarray}
\label{hamil}
{\tilde H} &=& H_c +\sum_{\alpha=1}^R (- \frac{1}{\lbrace\theta_\alpha,D_\alpha T_{\theta_\alpha}\rbrace}\lbrace H_c,C_\alpha T_\alpha\rbrace\;\theta_\alpha  + \frac {1}{2!} \left(\frac{1}{\lbrace\theta_\alpha,D_\alpha T_{\theta_\alpha}\rbrace}\right)^2\lbrace\lbrace H_c,C_\alpha T_\alpha\rbrace ,C_\alpha T_\alpha\rbrace \;\theta_\alpha^2\nonumber\\
&-& \frac {1}{3!} \left(\frac{1}{\lbrace\theta_\alpha,D_\alpha T_{\theta_\alpha}\rbrace}\right)^3\lbrace\lbrace\lbrace H_c,C_\alpha T_\alpha\rbrace ,C_\alpha T_\alpha\rbrace ,C_\alpha T_\alpha\rbrace \;\theta_\alpha^3+ \dots)\,,
\end{eqnarray}
where, again, the repeated index $\alpha$ inside of the Poisson brackets, in the expression above, does not mean a sum. Any way, this is not seem a problem because the expression of Eq.(\ref{hamil}) can be elegantly written in terms of a projection operator on $H_c$, given by
\begin{equation}
\label{hamil1}
{\tilde H} = P H_c\equiv H_c +\( \;\sum_{\alpha=1}^R \;(e^{-\frac{\theta_\alpha}{\lbrace\theta_\alpha,D_\alpha T_{\theta_\alpha}\rbrace}\,\hat\xi_\alpha}- 1): H_c\;\),
\end{equation}
where the operator $\,\hat\xi_\alpha\,$, applied to any function $\,\cal F\,$, is defined by the following relation
\begin{equation}
\hat\xi_\alpha \,{\cal F} \equiv \lbrace {\cal F}, C_\alpha T_\alpha\rbrace.
\end{equation}
It is interesting to point out that this situation also occurs in the usual constraint projector formalism \cite{Vyt}.

Finally, in order to eliminate the WZ auxiliary fields we must find a representation for each WZ variable written only in terms of the original phase space variable $(q_i,p_i)$, i.e.,\break $\theta_\alpha=f_\alpha(q_i,p_i)$. The algebraic form of this function is obtained 
imposing that it has the same infinitesimal gauge transformation displayed by $\theta_\alpha$, namely,

\begin{equation}
\label{rep}
\delta\theta_\alpha=\delta f_\alpha(q_i,p_i).
\end{equation}
Notice that a procedure of verifying if the resulting first class theory reproduces the same equation of motion or the spectrum(at a quantum level) of the initial second class model must be evaluated at the end of the application of the formalism. Thus, it is possible to derive a gauge-invariant Hamiltonian written only as a function of the original phase space variables $(q_i,p_i)$ satisfying the first class algebra

\begin{equation}
\label{fca}
\{ \tilde{H}, T_\alpha \} =0, \,\,\,\, \alpha=1,2,\dots, R.
\end{equation}

\section{Applications of the formalism}

\subsection{The reduced-SU(2) Skyrme model}

Few decades ago, Skyrme proposed to describe baryons as topological solutions of the SU(2) NLSM with an appropriate stabilizing term. 
The semi-classical quantization of the model was obtaining in \cite{ANW} separating the collective coordinate.  
Let us consider the SU(2) Skyrmion Lagrangian 

\be
\label{S10}
L = \int d^3x \left\{ \frac{f_\pi^2}{4} Tr\left(\partial_\mu U^\dagger \partial^\mu U\right) +\frac 1{32 e^2}
Tr\left[U^\dagger \partial^\mu U , U^\dagger \partial^\nu U\right]^2\right\},
\ee
where $f_\pi$ is the pion decay constant and $e$ is a dimensionless parameter. $U$ is a $SU(2)$ matrix transforming as $U\rightarrow AUB^{-1}$ under chiral $SU(2)\times SU(2)$, satisfying the boundary condition $\lim\limits_{r\rightarrow \infty} U= I$ so that the pion field vanishes as $r$ goes to infinity.  There are soliton solutions described by the action (\ref{S10}) whose topological number are identified with the baryon number. To describe the static soliton we start with the {\it ansatz} $U(r)=\exp\{i\vec\tau_a {\hat x}_a f(r)\}$ where $\vec\tau_a$ are Pauli matrices, ${\hat x} = {\vec x}/r$  and $\lim\limits_{r\rightarrow \infty} f(r) =0$ and $f(0)=\pi$. Performing the collective semi-classical expansion in (\ref{S10})\cite{ANW}, where $U(r,t)= A(t) U(r) A^\dagger(t)$ and $ A \in SU(2)$, we obtain after performing the space integral,

\be
\label{S20}
L = -M + {\cal I} \; Tr \left( \partial_0 A \partial_0 A^{-1}\right).
\ee
M and ${\cal I}$ are the soliton mass and the moment of inertia, respectively, which in the hedgehog {\it ansatz} are given by

\be
\label{S30}
M = 2\pi\!\!\int_0^\infty \!dr  r^2\!\left[  {f_\pi^2}\left(\left({\frac {df}{dr}}\right)^2 + 2 \frac{\sin^2{f}}{r^2}\right) \!+ \frac{\sin^2{f}}{e^2 r^2}\left( 2\left({\frac {df}{dr}}\right)^2 \!+ \frac{\sin^2{f}}{r^2}\right)\right],
\ee
and

\be
\label{S40}
{\cal I} = \frac {8\pi}{3} \int_0^\infty dr \;  r^2 \sin^2{f}\left[ f_\pi^2 + \frac {1}{e^2}\left(\left({\frac {df}{dr}}\right)^2 +  \frac{\sin^2{f}}{r^2}\right)\right] .
\ee
The unitary matrix $A$ may be represented by $A = a_0 + i\; {\vec a}.{\vec \tau}$, which satisfies the spherical constraint

\be
\label{S50}
a_ia_i - 1 = 0,
\ee
since the condition $AA^\dagger=1$ must be obeyed. In terms of these variables, the Skyrmion Lagrangian (\ref{S20}) becomes 

\be
\label{S60}
L = - M + 2\; {\cal I}\; \dot a_i\dot a_i + \zeta(a_ia_i - 1),
\ee
where $\zeta$ is a Lagrange multiplier that enforces the spherical constraint into the model. The corresponding Hamiltonian is

\be
\label{65}
H = M + \frac {1}{8{\cal I}} \pi_i\pi_i - \zeta(a_ia_i - 1),
\ee
where the canonical momenta conjugated to the collective coordinates $a_i$ are

\be
\label{66}
\pi_i = 4{\cal I} \dot a_i,
\ee
while the canonical momentum conjugated to the Lagrange multiplier $\zeta$ is, indeed, a primary constraint, which is read as

\be
\label{PR1}
T_1 = \pi_\zeta.
\ee
From the temporal stability condition, secondary, tertiary and quaternary constraints are required, namely,

\ba
\label{setcons}
T_2 &=& a_ia_i - 1,\nonumber\\
T_3 &=& a_i\pi_i,\\
T_4 &=& \frac {1}{8{\cal I}} \pi_i\pi_i + \zeta a_ia_i.\nonumber
\ea
Note that the last relation allows to fix the Lagrange multiplier, consequently, no more constraints arise from the iterative Dirac procedure. Due to this, the model has four second class constraints.

At this stage, we are ready to address the question of constraint conversion through the variational gauge-invariant formalism proposed in the last Section. The conversion process starts assuming that two second class constraints must be picked up to construct the gauge symmetry generators, and consequently two additional WZ variables, $\theta_1$ and $\theta_2$, are introduced. To put our work in perspective with other papers\cite{KHP,JCW}, we choose  the spherical constraint $T_2$, Eq.(\ref{setcons}) and the constraint $T_1$, Eq.(\ref{PR1}), to forge the first class constraints that will play the role of gauge symmetry generators. To this end, we begin to pick up the gauge symmetry generators as

\begin{eqnarray}
\label{TF}
{\tilde T}_2 &=& T_2 + \pi_{\theta_1}=a_ia_i-1+ \pi_{\theta_1},\nonumber\\
{\tilde T}_1 &=& T_1 + \pi_{\theta_2}=\pi_\zeta+\pi_{\theta_2}.
\end{eqnarray}
Recall that $\pi_{\theta_1}$ and $\pi_{\theta_2}$ are the WZ variables satisfying the canonical algebras 
$\lbrace \theta_1,\pi_{\theta_1}\rbrace= 1\,\,,\lbrace \theta_2,\pi_{\theta_2}\rbrace= 1\,\,$,  and ${\tilde T}_2$ and ${\tilde T}_1$ satisfy the first class algebra

\begin{equation}
\{ {\tilde T}_2, {\tilde T}_1 \}=0.
\end{equation}
The invariant Hamiltonian is written as

\begin{eqnarray}
\label{HF1}
\tilde{H} &=& H_c + G(a_i,\pi_i,\zeta,\theta_1,\theta_2)\nonumber\\
&=& M + {1\over 8 {\cal I} }\pi_i\pi_i - \zeta(a_ia_i - 1) + G^1_{(1)}\theta_1 + G^2_{(1)}\theta_2 + G^1_{(2)}(\theta_1)^2+ G^2_{(2)}(\theta_2)^2
+ \dots .
\end{eqnarray}

In agreement with the variational gauge-invariant formalism, the Hamiltonian ${\tilde H}$ must obey the variational principle $(\delta {\tilde H}^\alpha =\lbrace {\tilde H}, {\tilde T}_\alpha\rbrace = 0)$, i.e., this Hamiltonian must be invariant under the infinitesimal gauge transformations generated by symmetry generators ${\tilde T}_2$ and ${\tilde T}_1$.  

To compute all the corrections terms in order of $\theta_1$, we use the infinitesimal gauge transformations generated by symmetry generator ${\tilde T}_2$, given by

\ba
\label{IGTN}
\delta a_i &=& \varepsilon\{a_i,{\tilde T}_2\}=0,\nonumber\\
\delta \pi_i &=&\varepsilon\{\pi_i,{\tilde T}_2\}= -2\varepsilon a_i,\nonumber\\
\delta\zeta &=& \varepsilon\{\zeta,{\tilde T}_2\}=0,\\
\delta\theta_1 &=& \varepsilon\{\theta_1,{\tilde T}_2\}=\varepsilon,\nonumber\\
\delta\theta_2 &=& \varepsilon\{\theta_2,{\tilde T}_2\}=0,\nonumber
\ea
where  $\varepsilon$ is an infinitesimal parameter.

From the invariance condition $\delta{\tilde H}^\alpha =0$ (with $\alpha = 1, 2$)  given in Eq.(\ref{variational}), and using the infinitesimal gauge transformations (\ref{IGTN}), we can compute all correction terms in order of $\theta_1$. 
For linear correction term in order of $\theta_1$, Eq.(\ref{rdeltaH}), we get

\begin{eqnarray}
\label{IGT}
\delta H_c^1 +  G^1_{(1)} \delta {\theta_1} & = &  0,\nonumber\\
-\frac{1}{2{\cal I}}\varepsilon a_i\pi_i +  \varepsilon G^1_{(1)} &=& 0,\\
 G^1_{(1)} &=&  {1\over2{\cal I} } a_i\pi_i .\nonumber
\end{eqnarray}
For the quadratic term, Eq.(\ref{rdeltaH2}), we have

\ba
\delta G^1_{(1)} + 2  G^1_{(2)} \delta {\theta_1}  &=& 0,\nonumber\\
 -\frac {1}{{\cal I}}\varepsilon a_ia_i + 2 \varepsilon  G^1_{(2)} &=& 0,\\
G^1_{(2)} &=& {1\over 2{\cal I}} a_i a_i.\nonumber
\ea
For the tertiary term, we  obtain $ G^1_{(3)} = 0$, since $\delta G^1_{(2)}=\lbrace G^1_{(2)},{\tilde T}_2\rbrace = 0$. Due to this,  all correction terms $ G^1_{(n)}$ with $n\geq 3$ are null. 

To compute all the corrections terms in order of $\theta_2$, we use the infinitesimal gauge transformations generated by symmetry generator ${\tilde T}_1$ ,  given by

\ba
\label{IGTN2}
\delta a_i &=& \varepsilon\{a_i,{\tilde T}_1\}=0,\nonumber\\
\delta \pi_i &=&\varepsilon\{\pi_i,{\tilde T}_1\}= 0,\nonumber\\
\delta\zeta &=& \varepsilon\{\zeta,{\tilde T}_1\}=\varepsilon,\\
\delta\theta_1 &=& \varepsilon\{\theta_1,{\tilde T}_1\}=0,\nonumber\\
\delta\theta_2 &=& \varepsilon\{\theta_2,{\tilde T}_1\}=\varepsilon.\nonumber
\ea
For linear correction term in order of $\theta_2$, Eq.(\ref{rdeltaH}), we get

\begin{eqnarray}
\label{IGT2}
\delta H_c^2 + G^2_{(1)} \delta {\theta_2}  &=&  0,\nonumber\\
-\varepsilon (a_i a_i-1) +  \varepsilon G^2_{(1)} &=& 0,\\
 G^2_{(1)} &=&  a_i a_i -1 .\nonumber
\end{eqnarray}
For the quadratic term, Eq.(\ref{rdeltaH2}), we have

\ba
\delta G^2_{(1)} + 2 G^2_{(2)} \delta {\theta_2} &=& 0,\nonumber\\
G^2_{(2)} &=& 0.
\ea
Due to this result,  all correction terms $ G^2_{(n)}$ with $n\geq 2$ are null. 

Therefore, the gauge invariant Hamiltonian is

\be
\label{GIH}
\tilde{H} = M + \frac{1}{8 {\cal I}} \pi_i\pi_i - \zeta(a_i a_i - 1) 
+ {1\over2{\cal I} } a_i\pi_i\,\theta_1 + (a_ia_i - 1)\, \theta_2
+ {1\over 2{\cal I}} a_i a_i\, (\theta_1)^2.
\ee
This Hamiltonian, by construction, satisfies the gauge invariance property,

\ba
\lbrace{\tilde H},{\tilde T}_2 \rbrace = 0,\nonumber\\
\lbrace{\tilde H},{\tilde T}_1 \rbrace = 0.
\ea
If we fix the Wess-Zumino variables equal to zero, i.e., the unitary gauge, we recover the initial second class Skyrme model. 

The invariant Hamiltonian, Eq.(\ref{GIH}), can be elegantly written in terms of a gauge fields shifted,

\be
\tilde{H} = M + {1\over 2 {\cal I} } \tilde{\pi_i}\tilde{\pi_i} - {\tilde\zeta}(a_ia_i - 1),
\ee
where

\begin{eqnarray}
\label{shifts}
\tilde{\pi_i}&=&{\pi_i\over 2} + a_i \theta_1,\nonumber\\
{\tilde\zeta}&=&\zeta-\theta_2.
\end{eqnarray}
This algebraic expression reminds the field-shifting St\"uckelberg formalism\cite{stuckelberg}.

Here, it is interesting to point out that the variational gauge-invariant formalism was capable to solve exactly the problem proposed by Amorim, Barcelos and Wotzasek\cite{ABW} that is the complete embedding of the nonlinear sigma model\footnote{The BFFT 
formalism has solved this problem approximately.}. 
 
From the infinitesimal transformations generated by ${\tilde T}_2$ and ${\tilde T}_1$( Eqs.(\ref{TF})), ${\delta\theta}_1 = {\delta\theta}_2 = \varepsilon$, we can choose a representation for $\theta_1$ and $\theta_2$ as

\ba
\label{fix}
{\theta}_1 &=& - {a_i\pi_i\over 2 a_ja_j},\nonumber\\
{\theta}_2 &=& \zeta.
\ea
Substituting the relation above in the Eq.(\ref{GIH}), we get the invariant canonical Hamiltonian written only in terms of the original phase-space variables, given by 

\begin{eqnarray}
\label{HF3}
\tilde{H} &=& M + {1\over 8{\cal I}} \[ \pi_i\pi_i -  {(a_i\pi_i)^2\over a_ja_j}\],\nonumber\\
&=& M + \frac {1}{8{\cal I}}\pi_i M^{ij}\pi_j ,
\end{eqnarray}
where the phase space metric $M^{ij}$, given by

\be
M^{ij} = \delta^{ij} - \frac{a^ia^j}{a^2_l},
\ee
is a singular matrix which has $a_i$ as an eigenvector with null eigenvalue, namely,

\be
\label{singular}
a_iM^{ij} = 0.
\ee
Due to this, it is easy to show that the Hamiltonian (\ref{HF3}) is invariant under the infinitesimal gauge transformations (\ref{IGTN}) and (\ref{IGTN2}).  In view of this, the original second class constraints $T_2$ and $T_1$  become the gauge symmetry generators. 
Here, we can observe  the auxiliary tools characteristic of the WZ variables $\theta_1$ and $\theta_2$  because, at first, the WZ variables are introduced in the second class Hamiltonian with the purpose to enforce the symmetries. Next, they are replaced by an adequate representation leading to reveal the hidden symmetries present in the original phase-space variables.

Due to the fact that the matrix $\, M\,$ is singular, Eq.(\ref{singular}), at first, it is not possible to obtain the first class Skyrmion Lagrangian written only in terms of the original phase-space variables. For more details see Ref. \cite{TDL}.

In order to show the quantum equivalence of our first class Hamiltonian, Eq. (\ref{HF3}), and the initial second class Skyrme model, we will give an outline of the quantum mechanics treatment using for this the Dirac's first class procedure. A very detail description of our procedure can be found in reference\cite{JCW}. The physical wave functions must be annihilated by the first class operator constraints, reads as

\begin{eqnarray}
\label{qope}
T_2 |\psi \rangle_{phys} &= & 0,\nonumber\\
T_1 |\psi \rangle_{phys} &=& 0.
\end{eqnarray}

\noindent The physical states that satisfy (\ref{qope}) are

\begin{eqnarray}
\label{physical}
| \psi \rangle_{phys} = {1\over V } \, 
\delta(a_i a_i-1)\, \delta(\pi_\zeta)\, |polynomial\rangle.
\end{eqnarray}

\noindent where {\it V } is the normalization factor and $|polynomial \rangle ={1\over N(l)} (a_1+ i a_2)^l \,$. The corresponding quantum Hamiltonian is

\begin{eqnarray}
\label{echs1}
\tilde{H}= M+{1\over 8\lambda} \[ \pi_i\pi_i
-  {(a_i\pi_i)^2\over a_ja_j}\] .
\end{eqnarray}

\noindent The spectrum of the theory is determined by
taking the scalar product of the symmetrized invariant Hamiltonian, $_{phys}\langle\psi| \tilde{H} | \psi \rangle_{phys}\,$, given by

\begin{eqnarray}
\label{mes1}
_{phys}\langle\psi| \tilde{H} | \psi \rangle_{phys}=\nonumber \\
\langle polynomial |\,\,  {1\over V^2}  \int da_i\,d\pi_\zeta\,\,
\delta(a_i a_i - 1)\,\delta(\pi_\zeta)\,
\tilde{H}\,
\delta(a_i a_i - 1)\,\delta(\pi_\zeta)\,\,
| polynomial \rangle .
\end{eqnarray}
Integrating over $a_i$ and $\pi_\zeta$, we obtain 

\begin{eqnarray}
\label{mes2}
_{phys}\langle\psi| \tilde{H} | \psi \rangle_{phys}=\nonumber \\
\langle polynomial | M + {1\over 8\lambda}\[ \pi_i \pi_i 
- (a_i\pi_i)^2 \] | polynomial \rangle\nonumber\\= 
M+{1\over 8\lambda}\[ -\partial_j\partial_j+
{1\over2}\(OpOp+2Op+{5\over4} \)\]\nonumber\\
\label{EL}
= M+{1\over 8\lambda} \[ l(l+2)+{5\over 4} \],
\end{eqnarray}

\noindent where $Op$ is defined as $Op\equiv a_i\partial_i$. 
In Eq.(\ref{EL}),  the regularization of delta function squared  like $\delta^2 (a_i a_i - 1 )$  and $\delta^2 (\pi_\zeta)$ is performed by using the delta relation,  \break $ 2\pi\delta(0)=\lim_{k\rightarrow 0}\int dx \,e^{ik\cdot x} = \int dx = L.$ Then, we use the parameter L as the normalization factor.
It is important to point out that the energy levels, formula (\ref{EL}), is the same obtained in a constrained second class treatment of the SU(2) Skyrme model\cite{JAN}. Thus, this  result indicates that the variational gauge-invariant formalism produces a correct result when compared with the original second class system.

\subsection{Chiral boson quantum mechanics}

Chiral bosons, usually called self-dual fields in two space-time dimensions, have received much attention over the last decade because of their significant role played in the understanding of several models with intrinsic chirality, as heterotic strings\cite{gross} and quantum Hall effect\cite{stone}, for example. At the present time, it has experienced a revival since the study of the noncommutativity geometry became a relevant feature in the quantization of the Dp-brane in background $B_{\mu\nu}$ field\cite{jabari}. These models are specially interesting since the unique structure (Chern-Simons terms) that are available in three dimensions, give rise to topologically intricate phenomena without even-dimensions analogs. A quantum mechanical version of gauge field theory involving Chern-Simons terms has been proposed by Jackiw {\it et al.}\cite{DJT}. This was done in order to investigate in detail the change in symplectic structure that occurs when the vanishing of a parameter takes a second-order Lagrangian into a first-order one. 

The model proposed in Ref. \cite{DJT} is a quantum mechanical particle of mass {\it m} and charge {\it e} constrained to move on a two-dimensional plane, interacting with a constant magnetic field (B), which is orthogonal to the plane. This model has its dynamic governed by the following Lagrangian

\be
\label{L1}
 L = \frac{m}{2} {\dot q}_i^2 + \frac{B}{2}q_i\epsilon_{ij}\dot q_j - \frac{k}{2}q_i^2,
\ee
where $\epsilon^{ij}$ is an antisymmetric tensor, $(\epsilon^{12}= \epsilon_{12}=1)$.
This model is analogous to the Lagrangian density for the three-dimensional topological massive electrodynamic in the Weyl gauge $(A^0=0)$, reads as

\be
\label{L2}
{\cal L} = \frac{1}{2} \dot {\bf A}^2 + \frac{\mu}{2}{\bf A}  \times \dot {\bf A} - \frac{1}{2}(\nabla\times {\bf A})^2.
\ee
In Ref. \cite{DJT,bazeia} the behavior of the model (\ref{L1}) was investigated in the vanishing mass limit $(m\rightarrow 0)$ and was also shown to be analogue of a pure Chern-Simons(CS) gauge theory. Rescaling $A\rightarrow \sqrt{\frac{k}{\mu}}A$ and setting $\mu\rightarrow 0$, the Lagrangian above is reduced to the pure CS theory

\be
\label{L3}
{\cal L}_{CS} =  \frac{k}{2}{\bf A}  \times \dot {\bf A}.
\ee
Correspondingly, the vanishing mass limit in Eq.(\ref{L1}) produces the following Lagrangian

\begin{equation}
\label{LCSP}
L =  \frac{B}{2}q_i\epsilon_{ij}\dot q_j - \frac{k}{2}q_i^2,
\end{equation}
usually called Chern-Simons-Proca(CSP) quantum mechanical model. Recently, a similar approach was discussed in Ref. \cite{BS} in order to investigate the contribution of noncommutative geometry in the quantization of D3-brane in background {\bf B}-field.

The CSP model, described above, is an example of second class constrained theory since the constraints, given by

\begin{equation}
\label{scc}
T_i = p_i + {B\over 2} \epsilon_{ij} q_j \approx 0, \,\,\,\, (i=1,2)
\end{equation}
where $p_i={\partial L\over \partial \dot{q}_i}$ are the canonical momenta, satisfy the following Poisson algebra

\begin{equation}
\label{algebra}
\{ T_i, T_j \} = B\epsilon_{ij}.
\end{equation}
Due to this, the noncommutative nature of the model could be displayed after the computation of the Dirac brackets among the phase space coordinates. It will be done through the symplectic method\cite{FJ}. As this model is described by a first-order Lagrangian, given in (\ref{LCSP}), the symplectic variables $\zeta_\alpha$ and the respective one-form canonical momenta $A_{\zeta_\alpha}$ are

\ba
\label{variable}
\zeta_{q_i} &=&  q_i,\nonumber\\
A_{q_i} &=&  \frac{B}{2}q_i\epsilon_{ij},\nonumber
\ea
with the following Hamiltonian(symplectic potential)

\be
\label{HCSP}
H = {k\over 2} q_i q_i.
\ee
The corresponding symplectic matrix is

\be
\label{matrix}
f_{q_iq_j} =  \frac{B}{2}\epsilon_{ij}.
\ee
Since this matrix is nonsingular, it can be inverted to provide the noncommutative Dirac brackets, written as

\be
\lbrace q_i,q_j\rbrace_D =  \frac{2}{B}\epsilon_{ij}.
\ee
This complete our proposal of this section.

\subsection{The gauge invariant CSP model}

In this section, we are involved with the reformulation of CSP model as a gauge invariant theory by using the variational gauge-invariant formalism. The main feature behind of this formalism is the enlargement of the phase space with the introduction of an arbitrary function $G(q_i,p_i,\theta)$, given in Eq.(\ref{2060}), into the Hamiltonian. In agreement with this formalism, the CSP Hamiltonian, given in Eq. (\ref{HCSP}), becomes

\be
\label{HFCSP}
\tilde{H}= {k\over 2} q_iq_i + G(q_i,p_i,\theta).
\ee
Afterwards, an algebraic form is settle to be the symmetry generator which, for the present problem, we choose

\begin{equation}
\label{FCSP}
\tilde{T} = T_1 + p_\theta =  p_1 + \frac{B}{2}  q_2 + p_\theta,
\end{equation}
where  $p_\theta$ is a WZ variable which obeys the canonical algebra $\lbrace\theta,p_\theta\rbrace= 1$. Due to this, the infinitesimal gauge transformations are

\ba
\label{TC0}
\delta q_1 &=& \varepsilon\{q_1,\tilde T\}=\varepsilon,\nonumber\\
\delta p_1 &=& \varepsilon\{p_1,\tilde T\}=0,\nonumber\\
\delta q_2 &=& \varepsilon\{q_2,\tilde T\}=0,\\
\delta p_2 &=& \varepsilon\{p_2,\tilde T\}=- \varepsilon\frac{B}{2},\nonumber\\
\delta\theta &=&\varepsilon\{\theta,\tilde T\}= \varepsilon,\nonumber
\ea
where $\varepsilon$ is an infinitesimal parameter.

Following the variational gauge-invariant formulation sketched in Section II, the variational condition $\delta\tilde{H}=0$, given in Eq.(\ref{variational}), is

\be
\label{IGTC}
\delta\tilde{H} = \varepsilon\,\, \{H_c,\tilde{T} \} +  \sum_{n=1}^\infty \left(\delta G_{(n)}(q_i,p_i)\theta^n + n G_{(n)}(q_i,p_i)\theta^{(n-1)}\delta\theta\right) = 0.
\ee
From this general equation and using the relations, given in Eq.(\ref{TC0}), each correction term in order of $\theta$ can be computed. For the linear term in $\theta$, Eq.(\ref{rdeltaH}), we get 

\ba
\label{deltaHC}
\delta H_c+G_{(1)}\delta \theta &=& 0,\nonumber\\
\varepsilon k q_1 +  \varepsilon G_{(1)} &=& 0,\nonumber\\
 G_{(1)} &=& - k q_1.
\ea
For the quadratic term, Eq.(\ref{rdeltaH2}), we have

\ba
\label{deltaHC2}
\delta  G_{(1)} + 2 G_{(2)} \delta\theta &=& 0,\nonumber\\
-\varepsilon k + 2\varepsilon G_{(2)}  &=& 0,\nonumber\\
G_{(2)} &=&{k \over 2},
\end{eqnarray}
while for the tertiary term, we obtain

\ba
\delta G_{(2)} + 3 G_{(3)} \delta\theta &=& 0,\nonumber\\
 G_{(3)} &=& 0.
\ea
In view of this, all correction terms $ G_{(n)}$ for 
$n\geq 3$ are null. Thus, the invariant Hamiltonian with the WZ terms is

\begin{equation}
\label{HF2CSP}
\tilde H = {k\over 2} q_i q_i - k\,q_1 \, \theta + {k\over 2} \,\theta^2,
\end{equation}
which by construction satisfies the gauge invariance property

\be
\lbrace{\tilde H}, {\tilde  T }\rbrace = 0.
\ee
Note that the invariant Hamiltonian in Eq.(\ref{HF2CSP}) can be elegantly written in terms of a gauge field shifted, read as

\ba
\label{SHCSP}
\tilde H &=& {k\over 2} \[ (q_1 - \theta)^2 + q_2q_2 \]\nonumber\\&=& {k\over 2} \tilde{q_i}\tilde{q_i},
\ea
where

\ba
\tilde{q_1}&=& q_1 - \theta,\nonumber\\
\tilde{q_2}&=& q_2.
\ea
This algebraic expression reminds the field-shifting St\"uckelberg formalism\cite{stuckelberg}.

The main goal of the variational gauge-invariant formalism consists in to reveal the gauge symmetry existent in the model described by the original phase space fields. Since we consider the WZ variable as an auxiliary tool, we choose a representation for $\,\theta$ which preserves its infinitesimal gauge transformation given in Eq.(\ref{TC0}), reads as
 
\begin{equation}
\label{WZC}
\theta = - {2\over B}  p_2.
\end{equation}
Substituting the result (\ref{WZC}) in the Hamiltonian, Eq. (\ref{SHCSP}), we get a gauge invariant Hamiltonian written only in terms of the original phase space variables, reads as

\begin{eqnarray}
\label{HF3CSP} 
\tilde H &=& {k\over 2} q_iq_i + {2k\over B}q_1p_2
+ {2k\over B^2}p_2p_2 ,
\end{eqnarray}
with the constraint $\,\,T_1 = p_1 + {B\over 2} q_2\,\,$ as the gauge symmetry generator of the infinitesimal gauge transformations given in  Eq.(\ref{TC0}). It is easy to verify that $\tilde{H}$, given in Eq.(\ref{HF3CSP}), satisfies the first class algebra

\begin{equation}
\{\tilde H, T_1 \} = 0,
\end{equation}
and, consequently,  the Hamiltonian (\ref{HF3CSP}) is invariant under the infinitesimal gauge transformations (\ref{TC0}). 

As a result of our first class conversion formalism, since we have now Poisson brackets algebra, is the obtainment of  a commutative version of the Chern-Simons-Proca quantum mechanical  model.

The corresponding Lagrangian can be achieved by means of the constrained path integral formalism\cite{FS2}. The result is

\begin{eqnarray}
\label{Lagrange2}
\tilde{L} &=& {B^2\over 8k}\dot q_2 \dot q_2 - \frac{B}{2}\dot{q}_1 q_2-\frac{B}{2}q_1\dot{q}_2 - {1\over 2} k q_2q_2,\nonumber\\
&=&{B^2\over 8k}\dot q_2\dot q_2 - \frac{B}{2}\frac{d}{dt}({q}_1q_2) - {1\over 2} k q_2 q_2,
\end{eqnarray} 
being the corresponding action invariant under the infinitesimal transformations (\ref{TC0}). Note that $q_1$ variable has not appeared in Eq.(\ref{Lagrange2}) except into the total derivative. A similar result, using the BFFT formalism, was obtained in Ref. \cite{KIM}. From the Eq.(\ref{Lagrange2}), we obtain the following
equation of motion

\begin{equation}
\label{motion}
\ddot{q}_2=-{4k^2\over B^2} q_2,
\end{equation}
where it is just the usual harmonic oscillator having the frequency $\omega=2k/B$. Eq.(\ref{motion}) is the same obtained if we use the CSP second class Lagrangian, Eq.(\ref{LCSP}). This result  indicates the equivalence of our CSP gauge-invariant theory and the CSP second class model.

The invariant model given by the Lagrangian (\ref{Lagrange2}) presents the first class constraint $\,\,\tilde{T}_1=T_1 = p_1 + {B\over 2} q_2\,\,$
which is, according to the consistency of the formalism, the initial second class constraint  picked up  to forge the generator of the symmetry.

Finally, it is important to mention that the variational gauge-invariant formalism is capable to reveal the gauge symmetry existent on the original phase and configuration spaces, a result that was not yet discussed in the literature. 

\subsection{Chiral-bosons field theory}

Considerable attention has been given to the chiral-bosons field theory. This model is relevant to the comprehension of superstrings, W gravities, and general two-dimensional field theories in the light cone. Its apparent simplicity hides intriguing and interesting
points that remain until now.

The Floreanini-Jackiw (FJ) chiral-boson model has its dynamic governed by the following Lagrangian density\cite{FJ}

\begin{equation}
\label{CBTL}
{\cal{L}}= \dot{\phi} \phi^\prime-{\phi^\prime} ^2,
\end{equation}
where dots and primes represent derivatives with respect to time and space coordinates, respectively. Spacetime is assumed to be two dimensional Minkowskian variety. The primary constraint is

\begin{equation}
\label{CBTC}
T(\phi,\pi) = \pi - \phi^\prime,
\end{equation}
and the canonical Hamiltonian is

\begin{equation}
\label{CBTH}
{\cal{H}}_c= {\phi^\prime}^2.
\end{equation}

The time stability condition for the constraint $T$ does not lead to any new one because it satisfies the following Poisson bracket relation

\be
\label{CBC}
\lbrace T(x), T(y)\rbrace = - 2\delta'(x - y).
\ee

The goal of this section is to open up the possibility to implement a consistent covariant quantization of FJ chiral-boson. To this end, the variational gauge-invariant formalism will be used. This formalism begins introducing an arbitrary function into the Hamiltonian,

\be
\tilde {\cal H} = {\cal H}_c +  G(\phi,\pi_\phi,\theta)={\phi^\prime}^2 + G(\phi,\pi_\phi,\theta),
\ee
where $ G(\phi,\pi_\phi,\theta)$ is given by Eq.(\ref{2060}). 

The generator of gauge symmetry, $\,\tilde T\,$, is chosen as

\begin{equation}
\label{FCBT}
\tilde{T} = \pi - \phi^\prime + \theta,
\end{equation}

\noindent where the auxiliary field satisfies a non canonical Poisson bracket relation\footnote{It is not difficult to see that the constraint that leads to the brackets(\ref{NC}) is given by \\$T=\pi'_\theta+{1\over 2}\theta\approx 0$.}

\be
\label{NC}
\lbrace \theta(x), \theta(y)\rbrace = 2\delta'(x - y).
\ee
Combining Eqs.(\ref{CBC}) and (\ref{NC}), we have the first class Poisson bracket

\be
\label{algeb}
\lbrace \tilde T(x), \tilde T(y)\rbrace = 0.
\ee

In order to begin with the variational gauge-invariant formalism, the variational condition $(\delta\tilde {\cal H} = 0)$, given in Eq.(\ref{variational}) must be obeyed. The gauge infinitesimal transformations generated by $\tilde T$ are

\begin{eqnarray}
\label{DCBT}
\delta \phi(x) & = & \varepsilon\,\, \{ \phi(x),\tilde{T}(y)\} = \varepsilon \delta(x-y),\nonumber\\
\delta\pi(x) & = & \varepsilon\,\, \{  \pi(x),\tilde{T}(y) \} = -\varepsilon\delta'(x-y),\\
\delta {\cal H }_c(x) &=&
\varepsilon \{ { \phi'(x) }^2,\tilde{T}(y)\} =  2\,\varepsilon\,\phi'(x)\delta'(x-y),\nonumber\\
\delta \theta(x)&=& \varepsilon \,\{  \theta(x), \tilde{T}(y) \} = 2\,\varepsilon\delta'(x-y).\nonumber
\end{eqnarray}
Using these relations and following the prescription of the variational gauge-invariant formalism, the linear and quadratic correction terms are, respectively, obtained as 

\ba
\label{G1}   
\delta H_c + G_{(1)} \delta\theta &=& 0,\nonumber\\
 2 \varepsilon \phi'(x)\delta'(x-y) +  2 \varepsilon\delta'(x-y) G_{(1)} &=&0,\nonumber\\
G_{(1)} &=& -\phi',\\
\delta  G_{(1)} + 2 G_{(2)} \delta\theta &=& 0,\nonumber\\
-\varepsilon \delta' (x-y) + 4\varepsilon \delta' (x-y) G_{(2)} & = & 0,\nonumber\\
G_{(2)} &=& {1\over 4}.
\ea
As the secondary correction term is a scalar, all correction terms $G_{(n)}$, for $n \geq 3$, are null. Therefore, the gauge invariant Hamiltonian density is

\ba
\label{FHCBT2}
\tilde{\cal H} &=& {\phi^\prime}^2 - \phi'\theta + \frac 14 \theta^2,\nonumber\\
&=& (\phi' - \frac 12 \theta)^2.
\ea
Using Eqs.(\ref{DCBT}), it is easy to verify that $ \,\tilde{\cal{H}} $, Eq.(\ref{FHCBT2}), satisfies a first class algebra, given by

\be
\lbrace \tilde{\cal{H}}, \tilde T \rbrace = 0,
\ee
with $\tilde T = \pi - \phi^\prime + \theta$.

The gauge-invariant Hamiltonian, Eq.(\ref{FHCBT2}), is the same obtained by Amorim and Barcelos in \cite{AB} via BFFT 
formalism\footnote{Here, it is appropriate to comment that chiral-boson field theory is an example which the BFFT scheme does not necessarily involve fields canonically conjugated to the Wess-Zumino fields.} with the advantage that we have used few algebraic steps. Then, our results indicate the equivalence between the variational gauge-invariant formalism and the BFFT first class conversion method. 

The obtainment of the corresponding density Lagrangian is just a matter of direct calculation by means of the constrained path integral formalism
and the result is the same obtained in Ref.  \cite{AB}. It is opportune to comment that in the chiral bosons model, at first, is not possible to choose an adequate representation for the WZ field in terms of the original phase space variables. It occurs due to the singular property of the FJ chiral-boson model, whose constraint, Eq.(\ref{CBTC}), satisfies a second class algebra, given in Eq.(\ref{CBC}). Thus, it is necessary, in principle,  to adding a WZ variable  in the obtainment of the first class algebra, Eq.(\ref{algeb}).

\section{Conclusions}

In this paper, we have proposed a new approach to reformulate second class systems as gauge invariant theories. This gauge-invariant formalism is based on early conception that invariant models satisfy the variational principle. Following this idea and the Faddeev's suggestion\cite{FS}, we apply the variational principle on the WZ extended system in order to unveil symmetries present on the original second class system. One important feature of this formalism is the possibility to choose a convenient gauge symmetry generator which allows to investigate physical properties connected to gauge symmetries. In general, it is possible to substitute the WZ fields into the extended Hamiltonian that, subsequently, generates a gauge invariant Hamiltonian written in terms of the original phase-space fields. It is a meaningful characteristic displayed by the variational gauge-invariant formalism.
Another point that deserves to mention is the simple algebraic computation of the WZ extended Hamiltonian  when compared with other constraint conversion formalisms\cite{BFFT,embed,JW}. The variational gauge-invariant formalism was applied to different physical systems. First, we consider the reduced-SU(2) Skyrme model, where we have obtained a first class version for the original second class model. At this point, it is important to recall that the BFFT formalism has some additional difficulty to convert the whole set of second class constraints to first one\cite{ABW,BN}. Here, in this paper, we avoid this problem and, as a consequence, we have a success to convert the whole set of second class constraints to first one. We have also computed the energy spectrum, which reproduces the same results obtained in the literature\cite{KHP,JCW}. Second, we reformulate the CSP model as an invariant model which is written only in terms of the original phase space variables. This theory now becomes commutative. As long we known, it is a new result not yet presents in the literature. Third, the second class nature of the two dimensional self-dual model has been changed to the first one through the variational gauge-invariant formalism, which reproduces the results, given in Ref. \cite{AB}, in an effortless way.

\section{ Acknowledgments}
This work is supported in part by FAPEMIG and CNPq, Brazilian Research Agencies.  In particular, C. Neves would like to acknowledge the FAPEMIG.

\end{document}